\documentclass[aps,pre,twocolumn, superscriptaddress, showpacs,amsmath,amssymb,final,letterpaper]{revtex4}

\usepackage[utf8]{inputenc}
\usepackage{calc}
\usepackage[leftbars]{changebar}
\usepackage{graphicx}
\usepackage{amsmath,amssymb,amsthm}
\usepackage{multirow}
\usepackage{subfigure}

\usepackage{array}
\newcolumntype{L}[1]{>{\raggedright\let\newline\\\arraybackslash\hspace{0pt}}m{#1}}
\newcolumntype{C}[1]{>{\centering\let\newline\\\arraybackslash\hspace{0pt}}m{#1}}
\newcolumntype{R}[1]{>{\raggedleft\let\newline\\\arraybackslash\hspace{0pt}}m{#1}}

\usepackage{mathtools}

\usepackage{xcolor}

\newcommand{\BStructure}{\,^{(S)}B}
\newcommand{\BNode}{\,^{(N)}B}
\newcommand{\GStructure}{\,^{(S)}G}
\newcommand{\GNode}{\,^{(N)}G}
\newcommand{\XB}{\,^{(X)}B}
\newcommand{\XG}{\,^{(X)}G}
\newcommand{\SB}{\,^{(S)}B}
\newcommand{\SG}{\,^{(S)}G}
\newcommand{\NB}{\,^{(N)}B}
\newcommand{\NGown}{\,^{(N)}G}
\newcommand{\NSG}{\,^{(N,S)}G}
\newcommand{\NSB}{\,^{(N,S)}B}

\begin{document}

\author{Laurent H\'ebert-Dufresne}
\thanks{These authors contributed equally to this work.}
\affiliation{D\'epartement de Physique, de G\'enie Physique, et d'Optique, Universit\'e Laval, Qu\'ebec (Qu{\'e}bec), Canada G1V 0A6}
\affiliation{Santa Fe Institute, Santa Fe, NM, 87501}
\author{Edward Laurence}
\thanks{These authors contributed equally to this work.}
\affiliation{D\'epartement de Physique, de G\'enie Physique, et d'Optique, Universit\'e Laval, Qu\'ebec (Qu{\'e}bec), Canada G1V 0A6}

\author{Antoine Allard}
\affiliation{Departament de Física Fonamental, Universitat de Barcelona, Martí i Franquès 1, 08028 Barcelona, Spain}
\author{Jean-Gabriel Young}
\affiliation{D\'epartement de Physique, de G\'enie Physique, et d'Optique, Universit\'e Laval, Qu\'ebec (Qu{\'e}bec), Canada G1V 0A6}
\author{Louis J. Dub\'e}
\affiliation{D\'epartement de Physique, de G\'enie Physique, et d'Optique, Universit\'e Laval, Qu\'ebec (Qu{\'e}bec), Canada G1V 0A6}

\title{Complex networks as an emerging property of hierarchical preferential attachment}

\begin{abstract}
Real complex systems are not rigidly structured; no clear rules or blueprints exist for their construction. Yet, amidst their apparent randomness, complex structural properties universally emerge. We propose that an important class of complex systems can be modeled as an organization of many embedded levels (potentially infinite in number), all of them following the same universal growth principle known as preferential attachment. We give examples of such hierarchy in real systems, for instance in the pyramid of production entities of the film industry. More importantly, we show how real complex networks can be interpreted as a projection of our model, from which their scale independence, their clustering, their hierarchy, their fractality and their navigability naturally emerge. Our results suggest that complex networks, viewed as growing systems, can be quite simple, and that the apparent complexity of their structure is largely a reflection of their unobserved hierarchical nature.
\end{abstract}
\maketitle 

\section{Introduction}

The science of complexity is concerned with systems displaying emerging properties; systems where the properties of the whole do not directly follow from the properties of the parts \cite{Simon_hierarchy}. However, we intend to show how one property of the whole, \emph{hierarchy}, can alone be the origin of more complex features. We will describe hierarchical systems through a general model of \emph{colored balls in embedded bins} which itself explains the emergence of other features through the projection of these hierarchical systems onto complex networks. 

Most real networks tend to feature properties not found in most classic models of sparse random networks: \emph{scale-independence}, fat-tailed degree distribution \cite{Champernowne, Barabasi1999}; \emph{modularity}, the grouping of nodes in denser groups \cite{strogatz, Girvan, Hebert2011_prl}; \emph{hierarchy}, the embedding of multiple levels of organization \cite{ravasz, clauset}; \emph{fractality}, the self-similarity between levels of organization \cite{Song05_nat, Song06_natphys}; and \emph{navigability}, the possibility of efficient communication through a hidden metric space \cite{boguna_natphys, boguna_natcomm, papadopoulos}.

Sophisticated algorithms can be designed to reproduce most of these features, often based upon a multiplicative process to force their emergence by reproducing a basic unit on multiple scale of organization \cite{ravasz2, palla_pnas}. These models are useful as they can create realistic structures and test hypotheses about measured data. However, these constructions are not intended to provide any insights on the underlying mechanisms behind the growth of the system.
	
In contrast, generative models are quite successful at suggesting principles of organization leading to specific properties. For example, simple models exist to propose possible origins for scale-independence  \cite{Barabasi1999} or of the small-world effect \cite{strogatz}, but they fail to model the emergence of properties not included by design. Consequently, the identification of new universal properties requires the creation of new generative models. It is fair to say that a single unifying principle has yet to be proposed. 

In this paper, we aim to close the gap between complex deterministic algorithms and simple stochastic growth models. The hierarchical nature of networks suggests that the observed links between nodes are merely projections of higher structural units \cite{clauset, Hebert2011_prl, lhd12} (e.g. people create groups within cities in given countries). These subsystems will be our focus. We use one general assumption to design an equally general model of hierarchical systems: all embedded levels of organization follow preferential attachment.

On the one hand, our model can be seen as a generalization of classical preferential attachment models \cite{redner00, doro00, doro01, redner01, Hebert2011_prl, lhd12}. We can thus apply methods developed in this context by generalizing them to hierarchical systems. On the other hand, our model fills the gap to previous studies wishing to introduce non-trivial structural properties, such as clustering and centrality.  Past models manipulate the networks through local rules to add, remove or rewire links: for instance, triadic closure \cite{volz04, bianconi14} or copying mechanisms \cite{redner05, redner13}. We find that complex properties emerge more naturally when changing the focus of the model from the actual network and its properties to the hierarchical system that produces it.

We validate this model on the well documented dataset of production entities in the film industry (i.e. producers produce films within companies in given countries). We then study the structure of the projection of this system onto a complex network of co-productions between film producers. Interestingly, the resulting networks feature a scale-independent hierarchical organization, community structure, fractality and navigability.

The paper is structured as follows. In Sec.~\ref{sec:PA}, we provide a brief review of preferential attachment (PA), followed by an overview of structural preferential attachment (SPA) in Sec.~\ref{sec:SPA}. In Sec.~\ref{sec:HPA}, we generalize this organization principle to a family of processes generating hierarchical systems of embedded structural levels. A particular process of this family is then algorithmically described in~\ref{sec:model} and  mathematically studied in~\ref{subsec:math}. In Sec.~\ref{sec:nets}, we explain how complex networks can be obtained from this process by projecting a hierarchical system onto a chosen structural level. Finally, our conclusions are presented in Sec.~\ref{sec:conc} and
a few technical details are covered in two short appendixes. 

\section{\textbf{Preferential attachment (PA)}\label{sec:PA}} The preferential attachment principle is a ubiquitous \textit{rich-get-richer} mechanism modeling complex systems of all sorts \cite{Yule2, Gibrat, Simon, Price, Champernowne, Barabasi1999, Hebert2011_prl}. It implies that the likelihood for a given entity to be involved in a new activity is roughly proportional to its total past activities. For instance, an individual with 10 acquaintances in a social network is roughly 10 times more likely to gain a new connection than one with a single acquaintance. This does not imply causation; the individual does not necessarily gain a new connection because of its existing ones, but merely that its past is a good indicator of its future activity. This simple mechanism leads to a scale-independant distribution of the activity in question, modeling any system where the distribution of a resource among a population roughly follows a power-law distribution. Consequently, the number $N_s$ of individuals with a share $s$ ($\in \mathbb{N}$) of the resource scales as $s^{-\gamma}$, where $\gamma$ is called the scaling exponent. 

In practice, we consider a discrete time process where, during a time step $\Delta t=1$, a new element $i$ of share $s_i = 1$ is introduced within the system with rate $B$ (birth event) or the share $s_j$ of an existing element $j$ is increased to $s_j+1$ with rate $G$ (growth event). We can write a rate equation governing the distribution of individuals $N_s$ with a given share $s$:
\begin{align}
N_s(t + 1) & = N_s(t) + B \delta _{s,1} \nonumber \\ & + \frac{G}{\sum sN_s(t)} \left[\left(s-1\right)N_{s-1}(t) - s N_s(t)\right] \;
\label{master0}
\end{align}
where $\sum sN_s(t)$ is the sum of all shares (total resource) used to normalize the transition probabilities and which rapidly converges to $\left(B+G\right)t$. Consequently, we will hereafter use, $\sum_s sN_s(t)=(B+G)t$ interchangeably whenever they appear. Since $B$ is the birth rate, the evolution of the normalized distribution $\{\tilde{N}_s(t)\}$ can be obtained by replacing $N_s(t)$ by $tB\tilde{N}_s(t)$:
\begin{align}
\left(t+1\right)&B\tilde{N}_s(t+1) = tB\tilde{N}_s(t) +  B\delta _{s,1} \nonumber \\ & + \frac{GB}{B+G}\left[\left(s-1\right)\tilde{N}_{s-1}(t) - s \tilde{N}_s(t)\right] \; .
\label{eq:PA_Distribution}
\end{align}
Solving at statistical equilibrium, i.e. $\tilde{N}_s(t+1) = \tilde{N}_s(t) = \tilde{N}^*_s$, yields
\begin{equation}
\left(1+s\frac{G}{B+G}\right)\tilde{N}^*_s = \frac{G}{B+G}\left(s-1\right)\tilde{N}^*_{s-1} + \delta _{s,1}
\label{eq:PA_Asymp}
\end{equation}
or more directly for $s>1$
\begin{equation}
\tilde{N}^*_s = \dfrac{\prod _{m=1}^{s-1} \frac{G}{B+G} m}{\prod _{m=1}^s \left(1+ \frac{G}{B+G}m\right)} \; .
\label{eq:PA_Sol}
\end{equation}
Asymptotically for $s\rightarrow \infty$, this steady state can be shown to scale as a power-law
\begin{equation}
\lim _{t,s\rightarrow \infty} \tilde{N}^*_s(t) \propto s^{-\gamma} \quad \textrm{with } \gamma = 2 + \frac{B}{G} \; .
\label{gamma_gen}
\end{equation} 

\section{\textbf{Structural preferential attachment (SPA)}\label{sec:SPA}}

With the ongoing focus on the modularity of complex systems, e.g. the community structure of networks, it is essential to be able to consider structural properties of real systems within preferential attachment processes. In a recent study, we have introduced colored balls to represent individuals in social systems where unique individuals (unique colors) are a resource for communities (boxes) and vice versa \cite{Hebert2011_prl,lhd12}. This can be mapped to a process where colored balls are placed in boxes. Balls of the same color are meant to represent different activities of the same individual, just as different boxes represent different structures growing by receiving new balls. We have extended preferential attachment to structured systems: just as an individual involved in more social groups is more likely to join a new group, a larger social group is more likely to gain new members than a small one. We have coined the name, structural preferential attachment (SPA), to
describe this first level of extension of the PA principle. 

In SPA, the two important quantities are the membership of a given color --- i.e., the number of structures in which that color is found --- and the size of a given structure --- i.e., the number of balls it contains. They can be followed by the rate equation approach of Eq.~(\ref{master0}). In distinction to the previous section, we now have a first structural level, and our notation reflects this extension by an extra index on the associated quantities. In the case of memberships, the share of a ball is now the number $m$ of apparitions in different structures, whereas in the case of sizes, the share of a structure is now the number $n$ of balls it contains. Hence, in both cases, the total resource is given by the sum of all balls found in the system, regardless of their colors. We can thus write
\begin{align}
&N_{1,m}(t + 1) = N_{1,m}(t) + \BNode_1\delta _{m,1} \nonumber \\ & + \frac{\GNode_1}{\left(\BNode_1+\GNode_1\right)t} \left[\left(m-1\right)N_{1,m-1}(t) - m N_{1,m}(t)\right] \;
\label{masterSPA1}
\end{align}
for the number $N_{1,m}$ of different colors with memberships $m$ at the first structural level. $\BNode_1$ and $\GNode_1$ now represent the rates of introducing a new color (birth) or re-using an old one (growth), respectively. Similarly, the number of structures $S_{1,n}$ of size $n$ evolves as
\begin{align}
&S_{1,n}(t + 1) = S_{1,n}(t) + \BStructure_1 \delta _{n,1} \nonumber \\ & + \frac{\GStructure_1}{\left(\BStructure_1+\GStructure_1\right)t} \left[\left(n-1\right)S_{1,n-1}(t) - n S_{1,n}(t)\right] .\;
\label{masterSPA2}
\end{align}
Since Eqs.~(\ref{masterSPA1}-\ref{masterSPA2}) and Eq.~(\ref{master0}) are similar, the normalized distributions, $\{\tilde{N}_{1,s}(t) \}$ and 
$\{\tilde{S}_{1,s}(t) \}$, with $X_{1,s}(t) = t \XB\tilde{X}_{1,s} (t)$ (with $X= N$ or $S$), satisfy  equations of the form (\ref{eq:PA_Distribution}) 
whose stationary solutions reproduce Eq.~(\ref{eq:PA_Sol})
\begin{equation}
\label{eq:SPA_Sol_Asymp}
\tilde{X}^*_{1,s} = \dfrac{\prod _{m=1}^{s-1} \frac{\XG_1}{\XB_1 +\XG_1} m}{\prod _{m=1}^s \left(1+ \frac{\XG_1}{\XB_1+\XG_1}m\right)} \; ,
\end{equation}
from which we recover the scaling exponents directly:
\begin{subequations}
\label{eq:subExponent}
\begin{align}
\lim _{t,m\rightarrow \infty} \tilde{N}^*_{1,m}(t) \propto m^{-\gamma _{N,1}} \quad \textrm{with } \gamma _{N,1}= 2 + \frac{\BNode_1}{\GNode_1} \; ,\\
\lim _{t,n\rightarrow \infty} \tilde{S}^*_{1,n}(t) \propto n^{-\gamma _{S,1}} \quad \textrm{with } \gamma _{S,1}= 2 + \frac{\BStructure_1}{\GStructure_1} \; .
\end{align} 
\end{subequations}
In the context of social networks, this new process leads to a scale-independent community structure, where both the distribution of members per community and the distribution of communities per individual asymptotically follow a power-law organization. However considering that this organization is found in distributions of friends \cite{Barabasi1999}, of members in social groups \cite{Hebert2011_prl} and of city population \cite{Zipf}, it is natural to ask the following: How would a preferential attachment occurring on \textit{multiple} structural levels influence the created system? It is a popular idea that complexity frequently takes the form of hierarchy and that a hierarchical organization influences the property of the whole independently of the nature of its content \cite{Simon_hierarchy}. With the recent successes of preferential attachment models, we hereafter propose a generalization for hierarchical systems.

\begin{figure*}
\centering
\includegraphics[trim = 0mm 0mm 0mm 0mm, clip, width=1.0\linewidth]{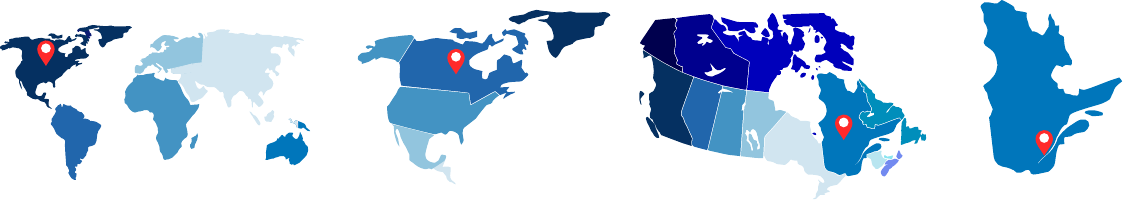}
\caption{(Color online) \textbf{An example of hierarchical structure.} Individuals involved in workshops taking place in given states or provinces forming countries within continents (four embedded structures).}
\label{fig:world}
\end{figure*}

\section{\textbf{Hierarchical preferential attachment (HPA)}\label{sec:HPA}}
We now generalize the process of Sec.~\ref{sec:SPA} by considering systems consisting of an arbitrary number $d$ of embedded levels of organization. Hence, we can describe Hierarchical Preferential Attachment (HPA) as a scheme of throwing colored balls in $d$ embedded levels of structures, which can be pictured as russian dolls but different.

\begin{figure*}
\includegraphics[width=0.95\textwidth]{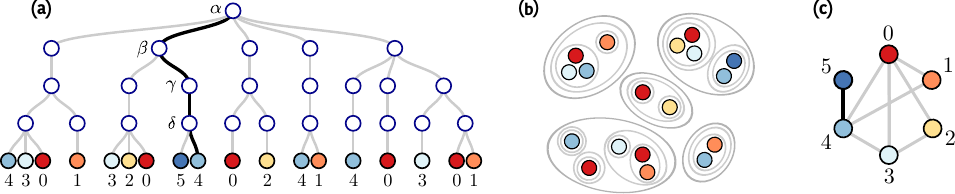}

\caption{(Color online) \textbf{Schematization of hierarchical preferential attachment.} HPA process frozen in  time as a ball labeled 4 (the label representing a ``color'') is added to a $d=3$ hierarchical structure. The process goes as follow. In this event, a structure at level 1 is chosen for growth (probability $1-p_1$). Among the 5 structures of level 1 (total size 8), the structure $\beta$ of size 2 is chosen for growth (probability $2/8$). Then, into the selected structure $\beta$, a smaller structure labeled $\gamma$ of size 1 is chosen for growth (probability $(1-p_2)\cdot 1/2$) and finally a level 3 structure labeled $\delta$ of size 1 (probability $(1-p_3)\cdot 1/1$). Since $q_{d=3}=1$ by construction, the ``color'' has to be new for $\delta$ (probability $q_3$). Then, the color is also new for $\gamma$ because it is a size 1 structure and the logical constraint applies. The color is chosen to be new for $\beta$ (probability $q_1$), but old for level 0 structure labeled $\alpha$ (probability $1-q_0$). At this point, the accessible ``colors'' are those labeled $1$ and $4$. Balls $0$, $2$, $3$ and $5$ can not be chosen since the color should be new for structure $\beta$. Balls $1$ and $4$ have the same probability of being chosen as they both belong to 3 level $1$ structures. The ball 4 is then chosen with probability $3/6$ and placed in $\delta$. (a) Hierarchical representation as an inverted tree. Navigating downwards corresponds to moving towards ever smaller structures until we reach the balls therein. (b) Representation as labeled balls in embedded levels of structures.
(c) Possible network representation of the system. In this case, two nodes share an edge if they belong to a same level 3 structure. Adding ball number 4 to structure $\delta$ creates the link highlighted in bold. }
\label{schema}
\end{figure*}

\begin{figure}
\centering
\includegraphics[trim = 0mm 0mm 0mm 0mm, clip, width=0.9\linewidth]{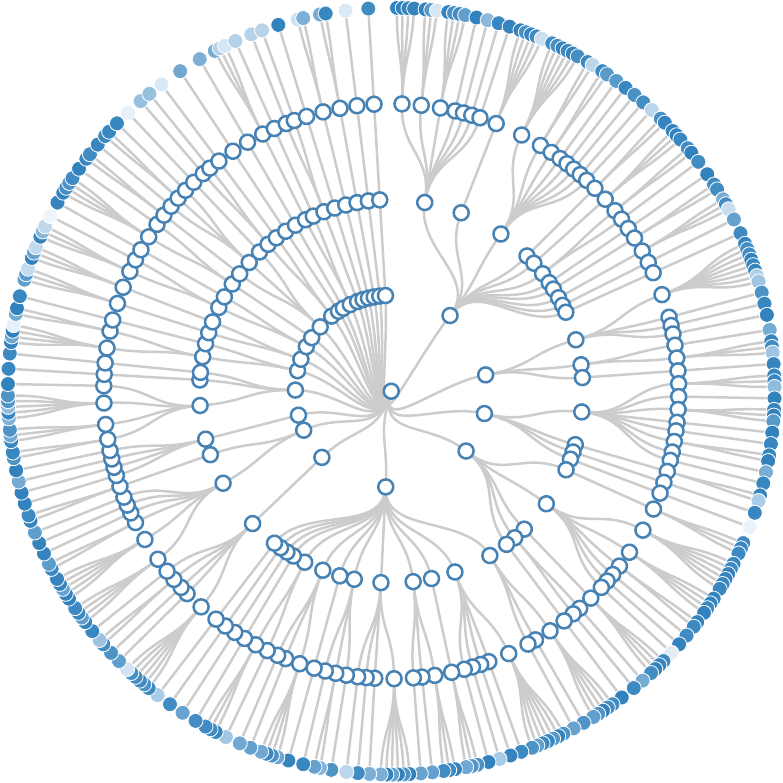}
\caption{(Color online) \textbf{An example of HPA.} HPA process for 250 steps on a structure of $d=3$ levels. Each radius represents a level of organization. 
The nodes are found at the outermost circle and a unique color (shades of blue, more than 50) specifies their identity. }
\label{fig:hpaExample}
\end{figure}

\subsection{\textbf{Qualitative description}\label{subsec:generalDescription}}
 
We will start with a tongue-in-cheek example of what a model of HPA could be. Obviously, preferential attachment is not meant to mimic the actual mechanisms or microscopic details at play in any system \cite{Simon}. In the following example, the proposed urn scheme should be seen as a potential abstract model for the system's statistical properties.

Let us assume that we want to study the distribution of scientists who have attended a small workshop held yearly in different states around the world. The network can then be constructed by ignoring time and simply assigning scientists to the different editions: we assign individual scientists to embedded structures (workshops held in geographical regions). Each structural level follows the preferential attachment principle based on the sub-structures they contain. Consider for instance the example of Fig. \ref{fig:world}: we are assigning one scientist to one workshop and to do so we must progressively go down the hierarchy of $d=4$ embedded structures. In this case, large-scale structures represent continents (level $k=1$) containing countries (level $k=2$) containing provinces or states (level $k=3$) containing fine-grained structures representing workshops (level $k=4$); this should be enough to describe the global system (planet Earth, level $k=0$). Thus, the level $k$ refers to different resolution of coarse-graining, such that large-scale structures mean low resolution (small $k$) and fine-grained structures mean high resolution (large $k$).

We now choose a workshop. On Fig. \ref{fig:world}, we associate the scientist to an existing continent --- $k=1$, in this case North America --- then within that continent we select a country --- $k=2$, Canada --- then a province or state --- $k=3$, Qu\'{e}bec --- and finally a single workshop --- $k=4$,  a workshop in Qu\'{e}bec City. At each level, the process follows the preferential attachment principle; e.g., the city was chosen proportionally to the number of workshops therein. 

We can now determine the identity of the scientist. This is achieved by lowering the resolution progressively and probing all structural levels with the following question: is the scientist a new participant? For instance, the scientist could be new to Canada (i.e. he has never attended a workshop in Canada), but not to North America, in which case his identity is borrowed from the United States or Mexico proportionally to his past activity in these two countries.

These embedded preferential attachment processes can be used to impose multiple constraints. Perhaps some countries host the workshop more often than others (preferential attachment at each structural level), and maybe some scientists seldom travel out of their own continent.

The HPA process can be mathematically described by using $d$ different versions of Eq.~(\ref{masterSPA1}) for the memberships of individuals (e.g., how many level $k$ structures in which a given individual is found) and $d$ more of Eq.~(\ref{masterSPA2}) for the sizes of structures (e.g., how many level $(k+1)$ structures in each level $k$ structures). The dynamics is then completely determined, assuming we set the birth 
$\NSB_k$ and the growth $\NSG_k$ rates properly at each level $k$.\\

\subsection{\textbf{Algorithmic description}\label{sec:model}}

 We now describe a particular HPA model based on Herbert Simon's preferential attachment process \cite{Simon} and explicitly show how it can be followed algorithmically. The next sub-section will then formalize the approach with an analytic description. Some visual representations of the model are given in Fig.~\ref{schema} and a large hierarchical structure simulated with HPA is presented in Fig~\ref{fig:hpaExample}. 

 The model is represented either as a literal system of balls in embedded bins, or as the hierarchy it describes. For the rest of the paper, we will refer interchangeably to a level $k$ structure as $k$--structure. Each event, or time step, is simply the act of throwing an additional ball in the system which depends on parameters $p_k$ with $k \in [0,d+1]$ and $q_k$ with $k \in [0,d]$. It will soon become clear that some of these parameters are trivially assigned to avoid irregularities in the equations: $p_0 = 0$, $p_{d+1}=1$, and $q_d =  1$.

\begin{table*}[ht!]
\begin{center}
\caption{\label{table1}\textbf{Notation}}
\begin{tabular}{l|l}
\toprule
$p_k$ & Probability to create a new $k$--structure.\\
$q_k$ & Probability to choose a new node for the selected $k$--structure.\\
$d$ & Number of structural levels of organization ($d=1$ for SPA).\\
\hline
$\BStructure_k$ & Rate of Structural Birth at level $k$.\\
$\GStructure_k$ & Rate of Structural Growth of a level $k$ structure (it implies the creation of a new structure at level $k+1$).\\
$\BNode_k$ & Rate of Nodal Birth at level $k$ (equivalent to the rate of adding a new node to the system).\\
$\GNode_k$ & Rate of Nodal membership Growth at level $k$ (rate at which a node acquires membership to an existing $k$--structure).\\
\hline
$S_{k,n}$ & Number of $k$--Structures of size $n$ (i.e., containing $n$ different $(k+1)$--structures).\\
$S_k$ & Number of Structures at level $k$ ($=\sum_n S_{k,n}$).\\
$N_{k,m}$ & Number of Nodes with $m$ memberships at level $k$ (i.e., appearing in $m$ different $k$--structures).\\
\hline
$P_k$ & Probability to choose a $k$--structure of size $1$ under PA.\\
$R_k(d)$ & Probability that the construction process ends by choosing an existing node at level $k$, considering $d$ levels of organization.\\
\toprule
\end{tabular}
\end{center}
\end{table*}

The general process goes as follows. At every time step $\Delta t=1$, an event takes place: a ball is thrown in $d$ embedded structures. We first choose a set of structures. Starting at level $k=1$, we have two options:
\begin{itemize}
\item[1a] With probability $p_k$, we create a new structure. This forces the creation of one structure at all deeper levels $k'>k$ within that new structure. A larger structure cannot exist without containing at least one smaller structure.
\item[1b.] With probability $1-p_k$, an existing $k$--structure is chosen for growth. It is done preferentially to its size, i.e., the number of $(k+1)$--structures that it contains. Repeat this step within the chosen structure (i.e. level $k+1$), until level $k=d$ is reached or until a structure is created at level $k<d$.
\end{itemize}
Once a new structure has been created at level $k$, or once the level $k=d$ is reached by choosing existing structures, the color of the ball must be determined. By construction, the ball is new for the $k$--structure. We must however determine if it is new for the containing ($k-1$)--structure. If not, we must sequentially examine lower resolution structures. This is determined by one of the two possible choices:
\begin{itemize}
	\item[2a.] With probability $q_k$, the color is new for the $k$--structure. We then move to the level $k-1$ and repeat the operation (2a or 2b). If level $k=1$ is eventually reached, a new color is introduced in the system and the two steps process ends.
	\item[2b.] With probability $1-q_k$, the color is chosen among all colors already occurring within this particular $k$--structure. This is done proportionally to the number of $(k+1)$--structures, embedded in that selected $k$--structure, in which the colors appear. The two steps process concludes.
\end{itemize}


\subsection{\textbf{Mathematical description}\label{subsec:math}}

The algorithmic rules just described can now be mapped onto an embedded system of preferential attachment equations. Table \ref{table1} 
gathers the different quantities involved. 

\subsubsection{Structural birth and growth}

The structures of level $k$ have a rate of birth, $\BStructure_k$, and of growth, $\GStructure_k$, for $k \leq d$, given by
\begin{subequations}
\label{Eq:lala}
\begin{align}
\BStructure_k  & =  \sum _{i=1}^{k}p_i\prod_{j=1}^{i-1}\left(1-p_j\right) \\ \shortintertext{and}
\quad \GStructure_k & = p_{k+1}\prod_{i=1}^{k}\left(1-p_i\right)
\end{align}
\end{subequations}
since birth events occur if structures are created at level $k$ or at a lower resolution ($k' < k$). The growth of a $k$--structure requires to choose existing structures at levels $1\leq i\leq k$ (probability $\prod_{i=1}^k(1-p_i)$) and the creation of a structure at level $k+1$ (probability $p_{k+1}$). In order to make every equation coherent, we adopt the product convention $\prod_{k=i}^j a_k=1$ and the sum convention $\sum_{k=i}^j a_k=0$ for $j<i$.

With these probabilities, the number $S_{k,n}(t)$ of $k$--structures with size $n$ can be approximately followed using Eq.~(\ref{masterSPA2}). However, while this is exact for the first structural level, it is an approximation for structures at a deeper level, $k > 1$. For example, the probability to choose a structure of size $n$ at level 2 will depend on the size $m$ of the level 1 structure in which it is nested. Mathematically, whereas level 1 evolves according to
\begin{align}
\nonumber S_{1,m}&(t+1) = S_{1,m}(t) +\BStructure_1\delta _{m,1} \\&+ \frac{\GStructure_1}{\left[\BStructure_1+\GStructure_1\right]t}\left[\left(m-1\right)S_{1,m-1}(t) - mS_{1,m}(t)\right] ,
\end{align}
 level 2 is governed by a somewhat more involved expression 
\begin{widetext}
\begin{align}
S_{2,n,m}(t+1) =  S_{2,n,m}(t)& + \frac{mS_{1,m}(t)}{\left[\BStructure_1+\GStructure_1\right]t} \left\lbrace \GStructure_2\frac{(n-1)S_{2,n-1,m}(t)-n S_{2,n,m}(t)}{\sum _i iS_{2,i,m}(t)} - \GStructure_1\frac{S_{2,n,m}(t)}{S_{1,m}(t)}\right\rbrace \nonumber \\
& \quad + \frac{(m-1)S_{1,m-1}(t)}{\left[\BStructure_1+\GStructure_1\right]t} \left\lbrace  \GStructure_1 \frac{S_{2,n,m-1}(t)}{S_{1,m-1}(t)} + \GStructure_1\delta _{n,1}\right\rbrace + \BStructure_1\delta _{n,1}\delta _{m,1}\label{allochien}
\end{align}
\end{widetext}
where $S_{2,n,m}(t)$ is the number of level 2 structures of size $n$, nested in a level 1 structure of size $m$. This equation takes into
account the choice of a 1--structure of size $m$ or $(m-1)$ and then the growth or the birth of a 2--structure. 

To reduce Eq.~({\ref{allochien}}) to a more manageable master equation of the form Eq.~(\ref{masterSPA2}), one must sum over all $m$ 
to obtain an equation for $S_{2,n}(t)= \sum_m S_{2,n,m}(t)$. 
Under the approximation 
\begin{eqnarray}
\langle n \rangle _{S,2,m} &\equiv& \sum _n n\frac{S_{2,n,m}}{mS_{1,m}}\nonumber\\
                                    &\simeq& \sum_n n \frac{S_{2,n}}{\sum_j jS_{1,j}}  = \sum_n n \frac{S_{2,n}}{\sum_j S_{2,j}}
                                        \equiv \langle n \rangle_{S,2}\nonumber \\
\end{eqnarray}
and using the relations $\sum_j jS_{k,j}=\left[\BStructure_k+\GStructure_k\right]t$ and $\SB_k + \SG_k = \SB_{k+1}$, the simplification follows immediately. This type of approximation can be applied successively to all levels (e.g. $\langle n \rangle_{S,3,i,j}=\langle n \rangle_{S,3}$), yielding equations similar to Eq.~\eqref{masterSPA2}. The stationary counterparts and scaling exponents 
(Eq.~\ref{eq:subExponent}) follow under the obvious replacements $(S_{1,n}, \SB_1, \SG_1) \to (S_{k,n}, \SB_k, \SG_k)$. 
 
 The resulting dynamical equations  therefore describe a set of {\em uncorrelated} levels of 
structural organization with  well-defined scaling exponents $\{\gamma_{S,k}\}$. 

\begin{figure*}[t!]
\centering
\includegraphics[trim = 0mm 0mm 0mm 0mm, clip, width=0.48\linewidth]{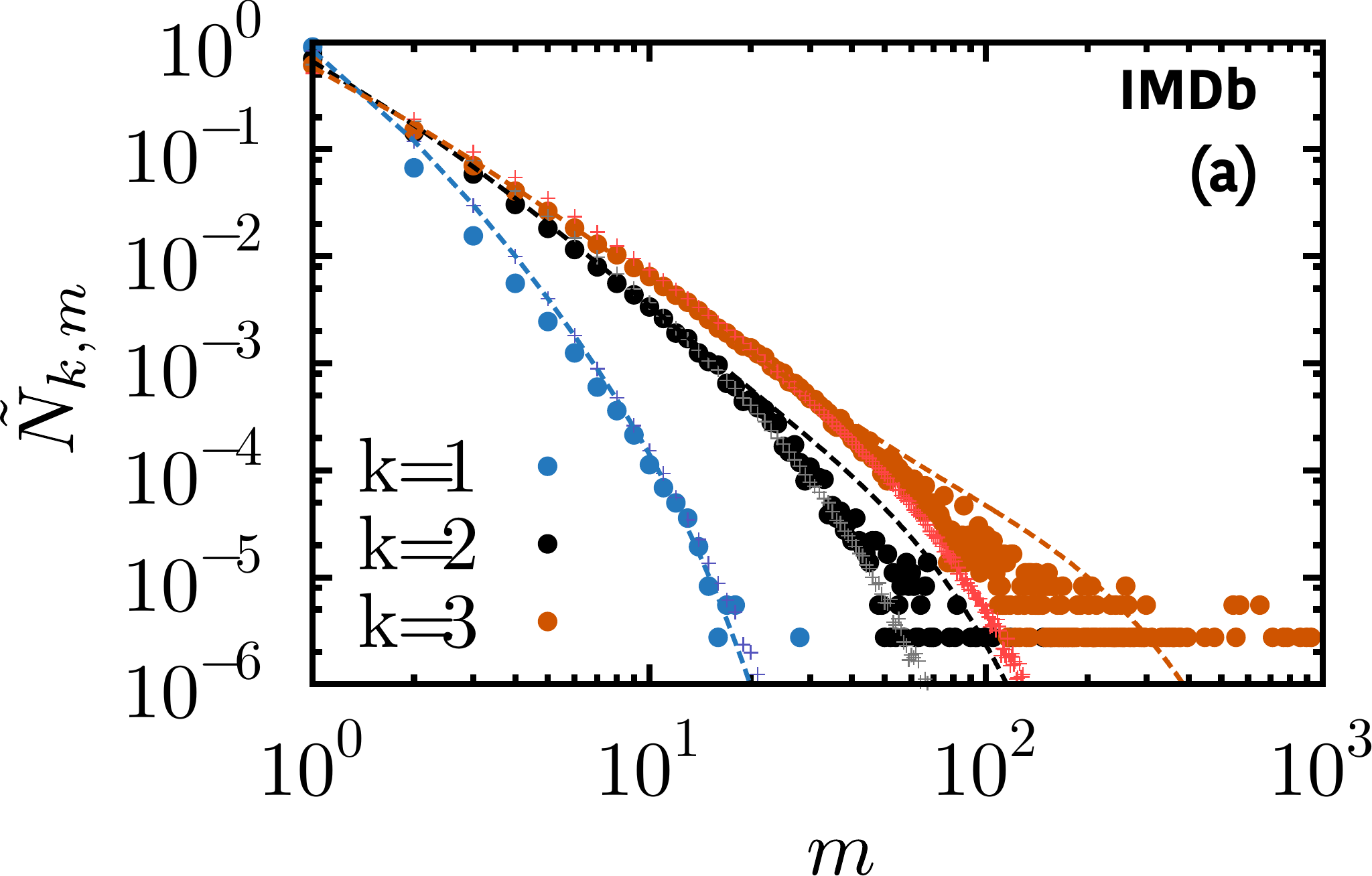}
\includegraphics[trim = 0mm 0mm 0mm 0mm, clip, width=0.48\linewidth]{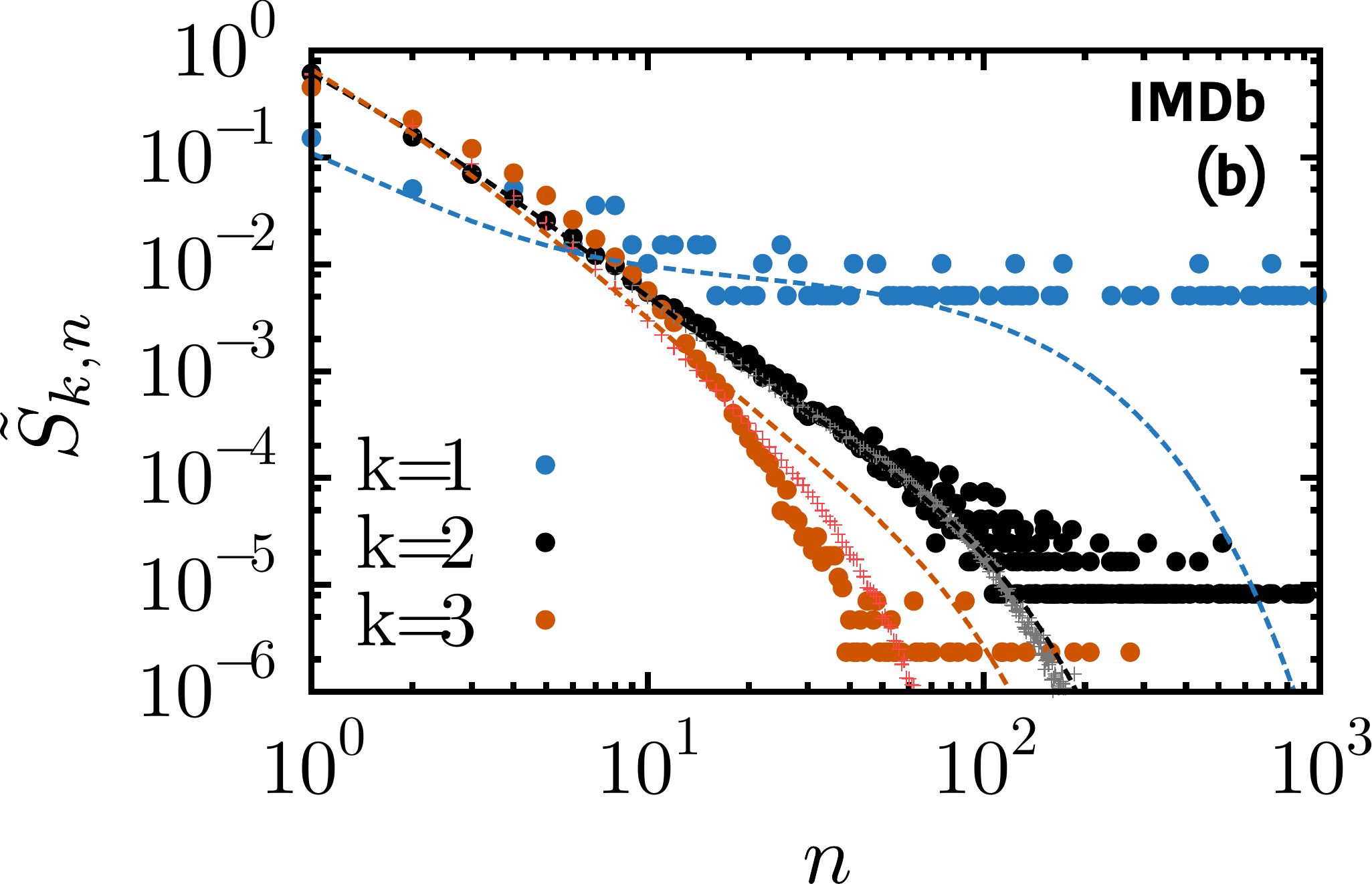}
\caption{(Color online) \textbf{Hierarchical structure of film production.} Events involving producers are distributed among $d=3$ structural levels: films at $k=3$ (upper dots in (a) and lower dots in (b)), in production companies at $k=2$ (middle dots in (a) and (b)) in countries at $k=1$ (lower dots in (a) and upper dots in (b)). (a) Distribution of the number of films/companies/countries a given producer is involved with. (b) Distribution of the number of producers/films/companies involved within a given film/company/country. The empirical data is shown with dots. Lines are obtained with Eqs~(\ref{masterSPA1}-\ref{masterSPA2}) using the corresponding birth and growth probabilities; crosses indicate direct Monte-Carlo simulations. Both calculations are iterated for $10^6$ time steps using $(p_1, p_2, p_3) = (0.0005, 0.185, 0.385)$, $(q_0, q_1, q_2) = (0.80, 0.60, 0.50)$. Simulated results of $S_{1,n}$ are not shown to avoid cluttering the figure (note that the plateau observed in the empirical data is due to finite size). The correspondence between the observed scale exponents and our mathematical results implies that the model is not over parametrized: $2d$ parameters for $2d$ scale exponents. The chosen parameters were hand selected to roughly reproduce the qualitative behavior of each distribution.}
\label{IMDb}
\end{figure*}

\subsubsection{Nodal birth and membership growth}

While the description of structure sizes is a straightforward problem, things get more involved for the number $N_{k,m}(t)$ of colors appearing in $m$ structures of level $k$. An important logical constraint occurs for $k$--structures with size equal to one: if the color is new for the sole structure of level $k+1$ therein (probability $q_{k+1}$), it must logically be new for the structure of level $k$ (as seen in the example of Fig.~\ref{schema}). Thus, the probabilities $\{q_k\}$ must be corrected to account for this logical constraint:
\begin{align}
q'_k(t) & = q_k + P_{k}(t)q_{k+1} \nonumber \\ & = q_k + \frac{\tilde{S}_{k,1}(t)}{\sum _n n \tilde{S}_{k,n}(t)}q_{k+1}  \label{EqIci}
\end{align}
where $P_{k}(t)$ is the probability that the $k$--structure of interest had a size equal to 1. In other words, if the color is new at level $k$, it can either be because of the initial probability $q_k$, or because it was forced to be new by the aforementioned logical constraint. Equation \eqref{EqIci} is only valid for $0<k<d$ since there is no correction at $k=0$ and $k=d$ ; $q_0=q'_0$ and $q_d=q'_d=1$.The probabilities $P_{k}(t)$ can be obtained from the master equation for sizes of $k$--structures (Eq.~\eqref{masterSPA2}), as well as from their steady state values in the limit $t\rightarrow \infty$. Together this yields
\begin{align}
\lim _{t\rightarrow \infty} q'_k(t) &= q_k + \frac{\tilde{S}^*_{k,1}}{\sum _n n\cdot \tilde{S}^*_{k,n}}q_{k+1}\\ 
          &= q_k + \frac{\BStructure_{k}+\GStructure_{k}}{\BStructure_{k}+2\GStructure_{k}}\frac{q_{k+1}}{ \langle n^* \rangle _{S,k}}
          \label{eq:qprime}
\end{align}
where the average size $\langle n^* \rangle _{S,k}=\sum_n n \tilde{S}^*_{k,n}$ corresponds intuitively to the ratio of the total rate to the birth rate:
\begin{equation}
\langle n^* \rangle _{S,k} = \frac{\left[\BStructure_{k}+\GStructure_{k}\right]}{\BStructure_{k}} .
\end{equation}
This result can also be obtained analytically and its demonstration is relegated to Appendix \ref{Appendix_AvrgSize}.
Inserting this last expression in Eq.~\eqref{eq:qprime} finally leads to 
\begin{equation}
\lim _{t\rightarrow \infty} q'_k(t) = q_k + \frac{q_{k+1}}{1+2\GStructure_{k}/\BStructure_{k}} \; .
\label{qcorr}
\end{equation}


\noindent It is then a matter of evaluating the birth $\BNode_k$ and growth rates $\GNode_k$ (see Table~\ref{table1}). 
To obtain the growth rates, let us consider the probability $R_k(d)$ that the chosen color is an existing one selected according to level $k$. These probabilities are easily calculated for the three deepest levels. By definition, $R_d(d)=0$, and :
\begin{align}
& R_{d-1}(d) = \left(1-q'_{d-1}\right)\prod _{i=0}^{d-1}\left(1-p_i\right) \\
 \shortintertext{and} & R_{d-2}(d) = \left(1-q'_{d-2}\right)p_{d-1}\prod _{i=0}^{d-2}\left(1-p_i\right) \nonumber \\ & \qquad\qquad + \left(1-q'_{d-2}\right)q'_{d-1}\prod _{i=0}^{d-1}\left(1-p_i\right) \; .
\end{align}
These probabilities yield a recursive expression for $k \leq d-2$:
\begin{align}
& R_{k}(d) = \left(1-q'_{k}\right)p_{k+1}\prod _{i=0}^{k}\left(1-p_i\right) \nonumber \\ & \qquad\qquad + \left(1-q'_{k}\right)q'_{k+1}\frac{R_{k+1}(d)}{1-q'_{k+1}}\; ,\label{eq:RRR}
\end{align}
starting from $R_{d-1}(d)$ given above. The terms $\GNode_k$ can then be written as the sum of the probabilities of choosing an existing node according to level $k$ or higher levels ($k'<k$):
\begin{equation}
\GNode_k = \sum _{i=0}^{k-1} R_{i}(d) \; .
\end{equation} 
To obtain the birth rate, we calculate the probability of introducing a new individual at each time step. Since an individual has at least one membership at each level, the birth rate at each level is namely the rate of introducing a new color to the system. In consequence, $\BNode_i=\BNode_j$ for all $i,j$. Using the normalization $\BNode_d(t) + \GNode_d(t) = 1$, since we always add at least one ball to a $d$--structure, we infer:
\begin{equation}
\BNode_k =  1 - \GNode_d = \dfrac{q_0}{1-q_0} R_0(d) .
\end{equation}

At this point, it is perhaps helpful to collect some of the explicit expressions of the birth and growth functions for a few hierarchical depths, say $d=1, 2,$ and 3.   Table (\ref{table2}) illustrates the construction scheme of these functions. A few observations are worth noticing. First, for internal consistency and as already used previously, one observes that
\begin{equation}
     \SB_k + \SG_k = \SB_{k+1} \quad , \quad  0 \leq k \leq d-1
\end{equation}
as clearly seen from the definitions \eqref{Eq:lala}.  Second, at level $d$, the birth and growth functions satisfy a normalisation condition for both structures and nodes ($X= S$ or $N$)
\begin{equation}
   \XB_d + \XG_d = 1 \ .
\end{equation}
 Finally, $\NB_k$ and $\NGown_k$ depend explicitly on the probabilities $\{q_k'\}$ which themselves depend on the structural functions $\{\SB_k\}$ and $\{\SG_k\}$. In other words, the logical constraints, captured by the  $\{q_k'\}$, induce correlations between the evolution of
the hierarchical structure and the distribution of elements within this structure.

\begin{table*}[ht!]
\begin{center}
\caption{\label{table2}\textbf{Birth and Growth Functions}}
\begin{tabular}{L{3cm}L{6cm}L{8cm}}
\toprule\\
$d=1$ (SPA)&$d=2$ & $d=3$\\\\
\hline\\
$\SB _0= 0$   & $\SB _0= 0$                & $\SB _0= 0$ \\
$\SB _1= p_1$ & $\SB _1= p_1$              & $\SB _1= p_1$\\
			  & $\SB_2= p_1 + p_2 (1-p_1)$ & $\SB_2= p_1 + p_2 (1-p_1)$\\
			  &							   & $\SB_3= p_1 + p_2 (1-p_1) + p_3 (1-p_1) (1-p_2)$\\\\

$\SG _0= p_1$      & $\SG _0= p_1$            & $\SG _0= p_1$ \\
$\SG _1= (1- p_1)$ & $\SG _1= p_2 (1- p_1)$   & $\SG _1= p_2 (1- p_1)$\\
				   & $\SG_2= (1-p_1) (1-p_2)$ & $\SG_2= p_3 (1-p_1) (1-p_2)$\\
   				   &      					  & $\SG_3=  (1-p_1) (1-p_2)(1-p_3)$\\\\

$R_0(1) = (1 -q_0)$ & $R_0(2)= (1-q_0) \left[ p_1 + q_1' (1 -p_1)\right]$ & $R_0(3)= (1-q_0) \left\{ p_1 + q_1' (1 -p_1)  \left[ p_2 + q_2' (1-p_2) \right] \right\}$  \\
					& $R_1(2) = (1-q_1') (1 - p_1)$                       & $R_1(3)= (1-q_1') (1-p_1) \left[ p_2 + q_2' (1 -p_2)\right]$\\
					&													  & $R_2(3) = (1-q_2') (1 - p_1)(1-p_2)$ \\\\

$\NB_1 = q_0$       & $\NB_{k} = q_0 \left[ p_1 + q_1' (1 -p_1)\right]$                & $\NB_{k} = q_0 \left\{ p_1 + q_1' (1 -p_1) \left[ p_2 + q_2' (1-p_2) \right]\right\}$\\
$\NGown_1 = R_0(1)$ & $\NGown_k=  \sum_{i=0}^{k-1} R_i(2)$  for $k=\lbrace 1,2\rbrace$ & $\NGown_k=  \sum_{i=0}^{k-1} R_i(3)$  for $k=\lbrace1,2,3\rbrace$\\\\
\toprule
\end{tabular}
\end{center}
\end{table*}

\begin{figure*}[t!]
\centering
\includegraphics[trim = 0mm 0mm 0mm 0mm, clip, width=0.32\linewidth]{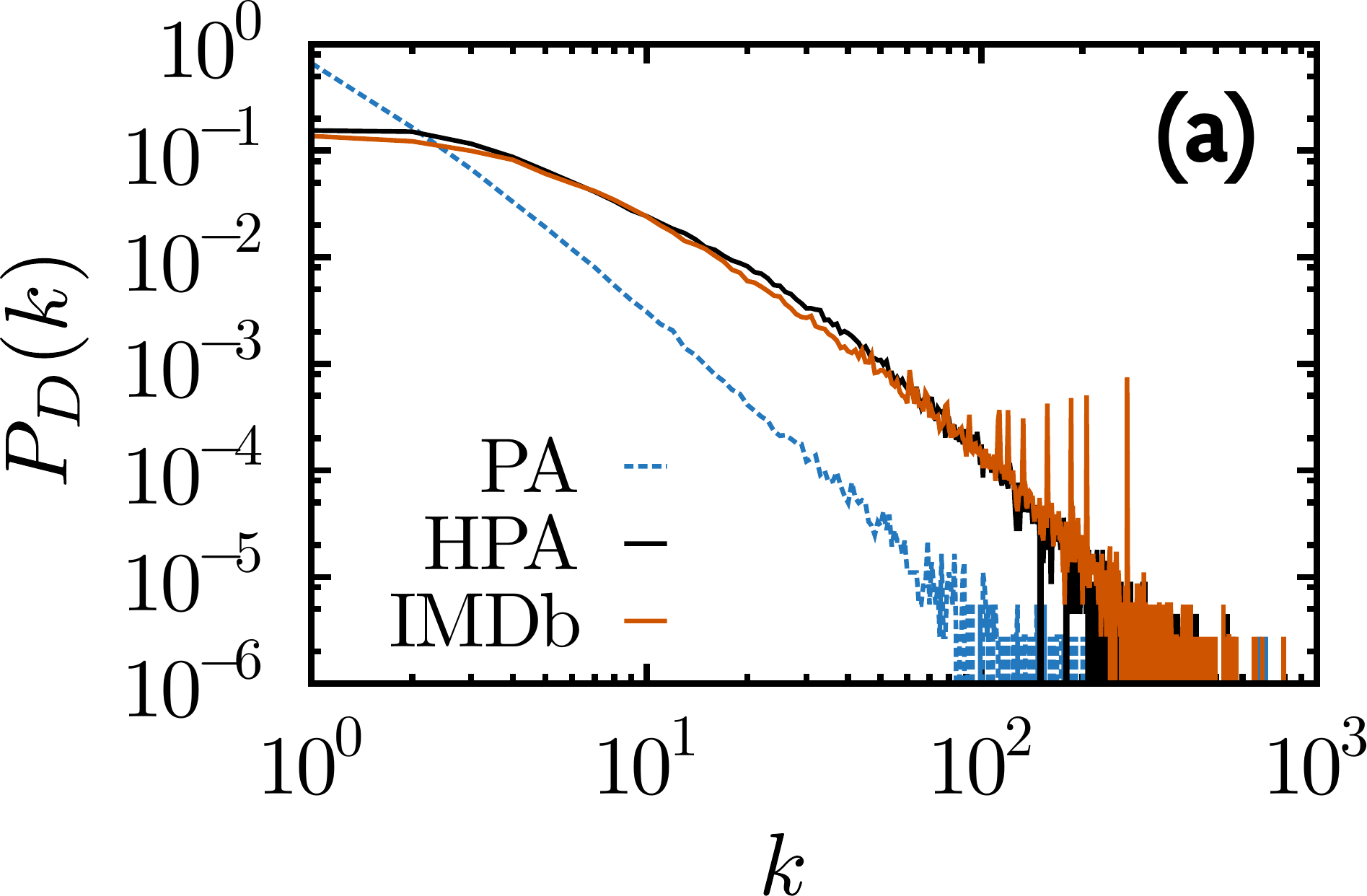}
\hspace*{2mm}
\includegraphics[trim = 0mm 0mm 0mm 0mm, clip, width=0.31\linewidth]{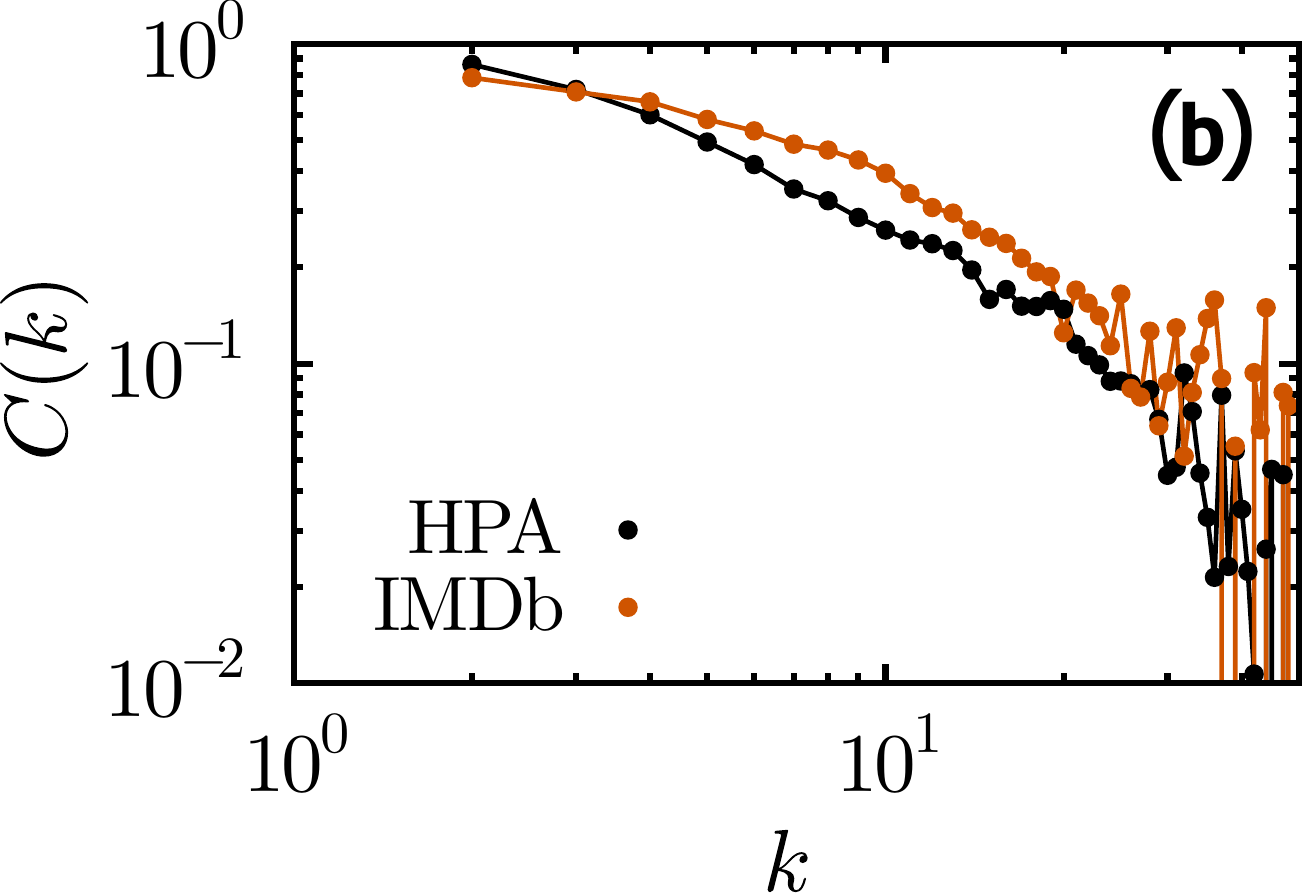}
\hspace*{2mm}
\includegraphics[trim = 0mm 0mm 0mm 0mm, clip, width=0.32\linewidth]{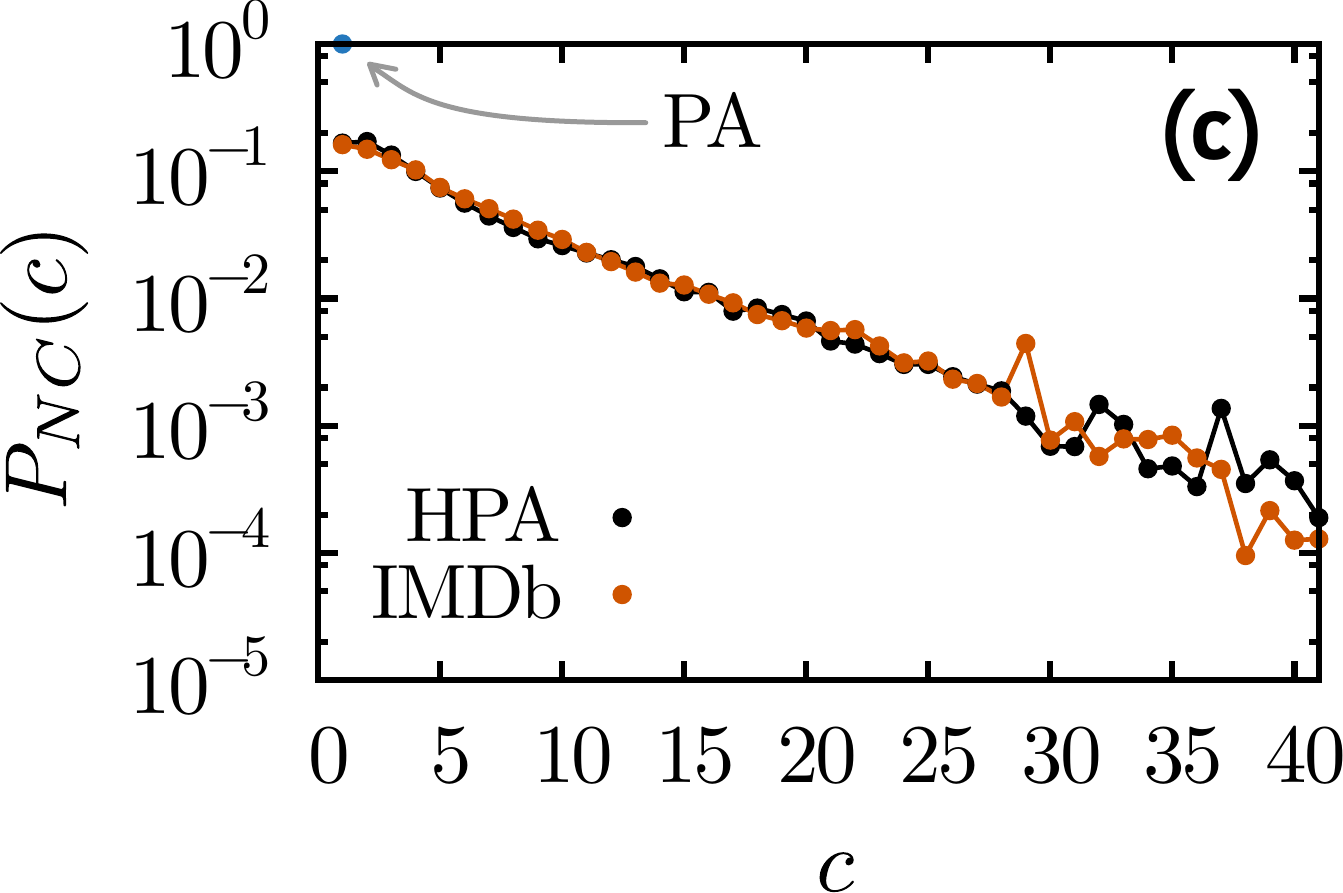}
\caption{ (Color online) \textbf{Scale-independence and clustering of a projected hierarchical systems.} (a) Degree distribution $P_D(k)$ observed in networks created by projecting the systems of Fig.~\ref{IMDb} on webs of co-productions (the actual data and one simulated system with the parameters used in Fig.~\ref{IMDb}). A network obtained through the classic preferential attachment model \cite{Barabasi1999} (PA) is given for comparison. (b) Average clustering coefficient for nodes of degree $k$.  PA leads to a vanishing clustering $C(k)=0$ for all degree $k$ in large networks. (c) Distribution of node centrality $P_{NC}(c)$ measured with their coreness $c$ under $k$-core decomposition of the networks. PA leads to a unique shell of coreness $c=1$ because of the tree-like structure
 of the network.}
\label{IMDb_prop}
\end{figure*}

To validate the use of these birth and growth rates in Eqs~(\ref{masterSPA1}-\ref{masterSPA2}), we examine the pyramid of production entities in the film industry. Based on the Internet Movie Database (IMDb), we study a system with $d=3$ structural levels where 363\,571 producers (colored balls) are assigned to films ($426\,913$ films, $k=3$) associated with one principal production company (121\,958 companies, $k=2$), in a given country (198 countries, $k=1$). The results of this case study are presented in Fig.~\ref{IMDb}. 

While the mean-field description of the distributions $\{ \tilde{N}_{k,m}(t) \}$ suffers from neglecting possible correlations from one resolution level $i$ to level $i+1$, the numerical 
simulations correctly reproduce the system and its finite size effects (distribution cut-off). 
The approximation of uncorrelated levels is also the source of the error observed in the mean-field description of the distributions
 $\{ \tilde{S}_{k,n}(t) \}$ for $k>1$ and becomes increasingly inadequate for larger $k$. The progression of error is essentially caused by the fact that a strict description of the third level, for instance, should not only be given in terms of $S_{3,n,m}(t)$ (see Eq.~\ref{allochien}), but of $S_{3,n,m,l}(t)$ describing the number of level 3 structures of size $n$ nested in level 2 structures of size $m$ themselves nested in level 1 structures of size $l$.

\section{\textbf{Projection on networks}\label{sec:nets}} Despite the advent of large datasets, few hierarchical systems are categorized and referenced as such. Consequently, research tends to focus on a single level of activity. For instance, the IMDb is often studied as a network of co-actors \cite{Barabasi1999,strogatz}, or as in Fig.~\ref{IMDb}, a network of co-productions where producers are connected if they have produced a film together (if they are found within a common level $d$ structure). Effectively, this implies that the system is reduced to a projection of all structural levels onto the chosen activity. While the involvement of actors and producers in films is well captured, their involvement in different companies and countries is somehow encoded, and more than often lost, in the resulting network. 

\subsection{\textbf{Degree, clustering, and centrality}}

Following the projection procedure schematically illustrated on Fig.~\ref{schema} (right), Fig.~\ref{IMDb_prop} presents some basic properties obtained by projecting the hierarchical system of film production onto a network of co-producers. We first investigate the degree distribution $P_D(k)$ (co-producing link per producer) and the clustering function $C(k)$ (probability that two links of a degree $ k$ producer are part of a triangle) of a network projection of a HPA system based on the parameters used in Fig.~\ref{IMDb}. 
The non-trivial clustering \cite{strogatz, ravasz} and the power-law tail of the degree distribution \cite{Barabasi1999}, properties ubiquitous in real networks, are reproduced in our framework as emergent features of the HPA model. Essentially, by only fitting the hierarchical structure of the IMDb co-production network, we get a good approximation of the complex properties of the network projection without having to directly account for them in the model. For an example of the calculation of the scaling exponents across multiple scales, we refer the reader to Appendix \ref{Appendix_MultipleScale}. 
   
 Moreover, Fig.~\ref{IMDb_prop} also presents the result of a centrality analysis known as core decomposition. This analysis relies on the concept of $c$-cores, i.e. the maximal subset where all nodes share $c$ links amongst one another. A node is assigned coreness $c$ if it belongs to the $c$-core but not to the $(c+1)$-core. This procedure effectively defines a periphery (low $c$) and core (high $c$) to the network and was recently shown to reflect structural organization beyond simple local correlations \cite{LHD13_HRN}. 
The HPA centrality distribution is seen to agree quite well with the data. This increases our confidence that the model effectively reproduces the structure of the real hierarchical system beyond the statistical properties previously considered in Fig.~\ref{IMDb}. 

\begin{figure*}[t!]
\centering
\includegraphics[trim = 0mm 0mm 0mm 0mm, clip, width=0.46\linewidth]{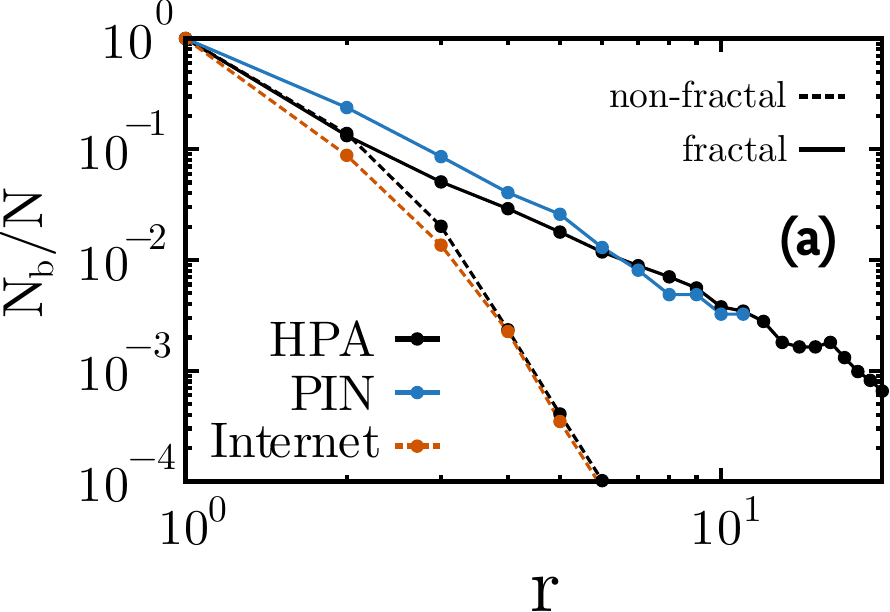}\qquad
\includegraphics[trim = 0mm 0mm 0mm 0mm, clip, width=0.46\linewidth]{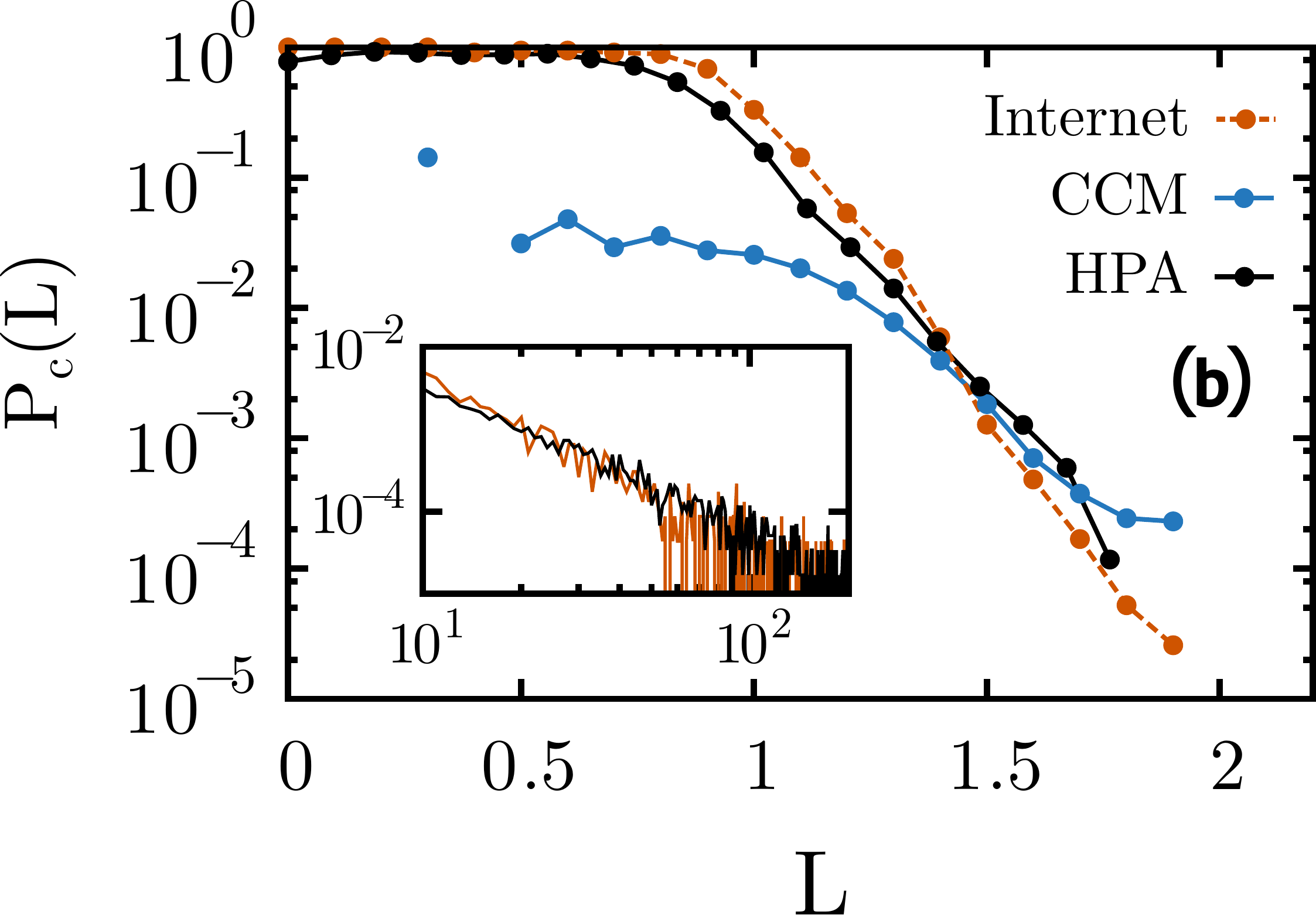}
\caption{ (Color online) \textbf{Fractality and navigability of projected hierarchical systems.} (a) Box counting results on a well-known fractal network (protein interaction network (PIN) of Homo Sapiens) and a non-fractal network (the Internet at the level of autonomous systems) \cite{Song05_nat}. HPA can approximately model how both of these networks span and cover their respective space, with $(p_1, p_2, p_3) = (0.01, 0.02, 0.30)$, 
$(q_0, q_1, q_2) = (0.95, 0.80, 0.30)$ (fractal) or $(p_1, p_2, p_3) = (0.005, 0.195, 0.395)$, $(q_0, q_1, q_2) = (0.60, 0.40, 0.30)$ (non-fractal). (b) Probability of connection $P_c(L)$ between nodes at a distance $L$ after an inferred projection of the networks onto an hyperbolic space. (The distance is given as a fraction of the hyperbolic disc radius. See Bogu\~{n}\'{a} \textit{et al.} \cite{boguna_natcomm} for details on the method.) Both the Internet and its HPA model are the same as presented on the left and share a similar scaling exponent for their degree distribution (see inset: degree distribution $D(k)$ versus $k$). The CCM corresponds to a rewired network preserving degree distribution and degree-degree correlations \cite{Newman02_PhysRevLett}, but  lacks the more complex structural correlations.}
\label{fractal}
\end{figure*}

\subsection{\textbf{Fractality}}

Aside from scale-independent degree distribution and non-trivial clustering function, the fractality of complex networks is often a sign of hierarchical organization \cite{Song05_nat, Song06_natphys}. One can unravel the fractal nature of a network using a box counting method: groups of nodes within a distance (number of links) $r$ of each other are assigned to the same box. The fractal dimension $d_b$ of a network manifests itself as a scaling relation between the number $N_b$ of boxes needed to cover all nodes and the size $r$ of the boxes ($N_b \propto r ^{-d_b}$). The self-similarity of network structure was previously assumed to stem from a repulsion or disassortativity between the most connected nodes \cite{Song06_natphys}. However, Fig.~\ref{fractal} demonstrates that fractality can also emerge from a scale-independent hierarchical structure, without further assumptions. Interestingly, Fig.~\ref{fractal}(left) also illustrates how, even if fractality might imply hierarchy, the opposite is not necessarily true.

HPA can produce both fractal and non-fractal networks. It remains to be determined whether or not this box counting method is truly equivalent to an actual measure of \emph{dimensionality}. However, it can, at the very least, be interpreted as an observation of how easily a network can be covered. 
Of course, since the definition of network fractality is somewhat ambiguous, so is the distinction between sets of HPA parameters leading to fractality or not. Nevertheless, a useful empirical rule can be established.

\begin{figure}[h!]
\centering
\includegraphics[width=0.93\linewidth]{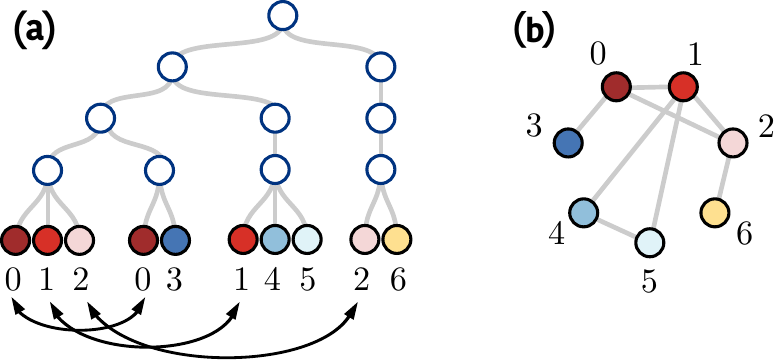}
\caption{(Color online) \label{Fracta2} \textbf{Example of bridges.} (a) Inverted tree representation of a hierarchical structure and (b) the corresponding network projection which shows how nodes labeled 0,1 and 2 act as bridges between structures.}
\end{figure}

Most models of stochastic network growth produce networks with very low mean shortest paths, low clustering and no long-range correlations. Consequently, the number of boxes needed to cover the whole network falls very rapidly. In HPA, we can control the manner in which boxes cover the network since the distance between higher structural levels is directly influenced by the memberships at this level. Hence, HPA can generate networks that are more robust to box covering (i.e. such that $N_b(r)$ falls slower with respect to $r$) if higher structural levels feature less nodes that act as bridges between structures and levels. For example, in  Fig.~\ref{Fracta2}, only nodes 0, 1 and 2 can be used by boxes to move from one level to the other (from workshops to countries, here illustrated as an inverted tree).

More precisely, let us consider the two different networks of Fig.~\ref{fractal} (left) built using the parameters given in the caption. Roughly speaking, in the non-fractal network, 2--structures contain on average around three 3--structures whereas nodes belong to over four 3--structures. Therefore, a single node typically grants access to all of the 3--structures contained within its 2--structure, such that a box covering at least \emph{part} of a 2--structure typically covers \emph{most} of it. The network is thus easily covered as higher levels are not any harder to navigate.

In contrast, 2--structures of the fractal network contain on average ten 3--structures while an average node may still be found within around three 3--structures. An average 2--structure may thus have nodes at a distance greater than three steps. The network is consequently harder to cover and can be expected to be much more robust to box-covering. As a general rule, we have found that to display \emph{measurable} network self-similarity, the average size of a structure (at level $k$) has to be at least greater than the memberships of a node at the deeper level (at level $k+1$).

\subsection{\textbf{Navigability}}

The box decomposition method tells us something about how networks cover the space in which they are embedded, and consequently at what speed a random walker might encounter new nodes in this network. However, it tells us nothing about the geometrical space that supports the network, or how a walker could find one specific node. In that respect, the navigability of complex networks has recently been a subject of interest for two reasons. First, the development of a mapping of networks to a geometrical space allows to predict the probability of links as a function of geometrical distance between nodes, which in turn enables an efficient navigation through the network \cite{boguna_natphys, boguna_natcomm}. Second, network growth based on preferential attachment fails to capture this geometrical property \cite{papadopoulos}. In a recent paper \cite{papadopoulos}, this metric was consequently considered as evidence of an opposition between two organizational forces: popularity (preferential attachment) and similarity (assortativity). Our last case study, shown in Fig.~\ref{fractal} (right), indicates that geometrical constraints, or network navigability, can emerge under a strict preferential attachment; which implies a growth driven by popularity only, but one occurring on multiple structural levels. The different hierarchical levels can \textit{a posteriori} be interpreted as indicators of similarity, but are conceptually much more general. 

We also compare in Fig.~\ref{fractal} (right) the results obtained on the actual network and on its HPA model with those obtained on a rewired network that preserves the degree distribution and degree-degree correlations (Correlated Configuration Model, CCM) \cite{Newman02_PhysRevLett}. The fact that 
CCM does not preserve the navigability of the Internet structure indicates that it emerges mostly from clustering and long-range correlations. As the HPA network does reproduce the navigability of the Internet, these long-range correlations could very well be consequences of the hierarchical structure. It would be instructive to investigate whether the inferred structure corresponds to the actual hierarchy of the Internet (probably of geographical nature: continents, countries, regions).


\section{\textbf{Conclusion}\label{sec:conc}} 

We have presented a proof of concept for a Hierarchical Preferential Attachment (HPA) model  in an attempt to reproduce the hierarchical nature of complex systems. We have illustrated  how complex networks could be better analysed by first modelling their hierarchical structure, and then projecting this structure onto a network. Not only does this procedure yields the non-trivial clustering of networks and their degree/centrality distributions at multiple levels, but it also gives access to the hidden geometrical metrics of these networks and the way they occupy space.

The fact that so many key features of the network structure are modelled using two minimal assumptions, hierarchy and preferential attachment, indicates that HPA provides more than theoretical insights; it leads support to the underlying assumptions. HPA could therefore be used to infer the possible hierarchical structure of networks when this information is not directly available.

Finally, while HPA is essentially a simple stochastic growth process, it nevertheless exemplifies eloquently how complex structural features of real networks --- e.g. scale-independence, clustering, self-similarity, fractality and navigability --- can emerge through the hierarchical embedding of scale independent levels. Perhaps this is the most important message: to study the structure of complex networks, one should avoid focusing on unique level of activity (e.g. links), but instead investigate the hidden hierarchical organizations from which the networks emerge.

\begin{acknowledgments}
The authors would like to acknowledge Calcul Qu\'{e}bec for computing facilities, as well as the financial support of the Canadian Institutes of Health Research (CIHR),  the Natural Sciences and Engineering Research Council of Canada (NSERC), the Fonds de recherche du Qu\'ebec--Nature 
 et technologies (FRQ-NT) and the James S. McDonnell Foundation. 
\end{acknowledgments}

\appendix
\section{Average structural size} \label{Appendix_AvrgSize}
We wish to demonstrate the following relation
\begin{equation}
	\langle n^* \rangle_{S,k}=\sum_{n=1}^{\infty}n \tilde{S}^*_{k,n}=\dfrac{\BStructure_k+\GStructure_k}{\BStructure_k}
\end{equation}
for the average size of $k$--structures at equilibrium. We will support our intuition that the mean value $\langle n^*\rangle_{S,k}$ should simply be the ratio between the number of shares $(\BStructure_k+\GStructure_k)t$ and the number of structures $\BStructure_kt$.
Inserting (\ref{eq:SPA_Sol_Asymp}) for $\tilde{S}_{k,n}$, simplifying  and rearranging, one finds 
\begin{equation}
	\sum_{n=1}^{\infty}n \tilde{S}^*_{k,n}=  b \sum_{n=1}^{\infty}\prod_{m=1}^{n}\dfrac{m}{b+m}
\end{equation}
where $b =(\BStructure_k+\GStructure_k)/\GStructure_k$. The numerator and denominator are easily identified. The numerator $\prod_{m=1}^n m= n!$ is a factorial while the numerator $\prod_{m=1}^s (b+m)=(b +1)_s$, is a Pochhammer symbol, i.e. $(x)_n= x(x+1)...(x+n-1)$. This reduces our expression to
\begin{equation}
	\langle n^* \rangle_{S,k} = b\sum_{m=1}^{\infty}\dfrac{s!}{(b+1)_s} .
\end{equation}
One recognizes the sum as a hypergeometric series (minus the leading term) namely
\begin{eqnarray}
	b\sum_{m=1}^{\infty}\dfrac{s!}{(b+1)_s} &=&b\sum_{m=1}^{\infty}\dfrac{(1)_s (1)_s }{(b+1)_s}  \frac{1}{s!} \nonumber \\
                                                                      &=& b\ [_2F_1(1,1;b+1;1) -1]\label{eq:2F1} .
\end{eqnarray}
Since the argument of the $_2F_1$ is equal to 1, a useful transformation \cite{GR} asserts that
\begin{equation}
	_2F_1(\alpha,\beta;\gamma;1)=_2F_1(-\alpha,-\beta;\gamma-\alpha-\beta;1)
\end{equation}
as long as Re $(\gamma) >\text{Re}(\alpha+\beta)$. This property applies to our case where $\alpha=\beta=1$ and $\gamma=b+1$ leading to a finite terminating series
\begin{equation}
	_2F_1(-1,-1;b-1;1)= 1 +\dfrac{(-1)(-1)}{(b-1)}
\end{equation}
which, once inserted in~\eqref{eq:2F1}, leaves us with the final result
\begin{equation}
	\langle n^* \rangle_{S,k}=b\left[\dfrac{1}{(b-1)}\right]=\dfrac{\BStructure_k+\GStructure_k}{\BStructure_k} .
\end{equation}

\section{Multiple scale independence}\label{Appendix_MultipleScale}
By ignoring the inter-level correlations for the structural growth, we have obtained in sub-section (\ref{subsec:math}) a set of coupled equations 
(\ref{masterSPA1}-\ref{masterSPA2}) for all levels $k$ that enable us to follow approximately the time evolution of the size distributions $\{\tilde{S}_{k,n}\}$
 and   of the node membership distributions $\{\tilde{N}_{k,m}\}$. We were then able to derive their scale exponents $\{\gamma_{S,k}, \gamma_{N,k}\}$ in the limit $t\rightarrow \infty$, Eq.~(\ref{eq:subExponent}). 

When investigating the projected properties of a hierarchical system, for instance the degree distribution of the resulting network, we can combine the membership and size distributions of the lowest level $d$ (where links are created) to deduce the resulting scaling exponent. As done in \cite{lhd12}, the idea is to define the following \textit{probability generating functions} (pgf):
\begin{align}
\mathcal{S}(x,t) = \sum _n \tilde{S}_{d,n}(t)x^n \;  \textrm{ and } \; \mathcal{N}(x,t) = \sum _m \tilde{N}_{d,m}(t)x^m \; .
\end{align}
Since a community of size $n$ implies $n-1$ links for each node, the generating function of the distribution of the number of links 
$\mathcal{L}(x,t)$ in a $d$--structure for a randomly chosen node can be generated by
\begin{equation}
\mathcal{L}(x,t) = \dfrac{\frac{d}{dx}\mathcal{S}(x,t)}{\frac{d}{dx}\mathcal{S}(x,t)\vert _{x=1}} 
                           = \dfrac{\sum _n n S_{d,n}(t)x^{n-1}}{\sum _n n S_{d,n}(t)}\; .
\end{equation}
The degree distribution  is then generated by $\mathcal{D}(x,t)$,  a pgf combining the distribution of memberships $m$ and of links obtained from each of these memberships:
\begin{equation}
\mathcal{D}(x,t) = \mathcal{N}(\mathcal{L}(x,t),t) \; ,
\end{equation}
which will simply scale as the slowest decreasing function between $\mathcal{N}(x,t)$ and $\mathcal{L}(x,t)$. The scale exponent of the degree distribution is thus given by
\begin{equation}
\textrm{min}\left[\gamma_{N,d}, \gamma_{S,d} -1\right]  \; .
\end{equation}
The same method could of course be used to determine the scaling of other projections (e.g., network of companies sharing or having shared at least one producer).


\begin{thebibliography}{10}

\bibitem{Simon_hierarchy}
H.~A. Simon,  {\em Proceedings of the
  American Philosophical Society} \textbf{106}, 467 (1962).

\bibitem{Champernowne}
D.~G. Champernowne,  {\em Economic Journal} \textbf{63}, 318 (1953).

\bibitem{Barabasi1999}
A.~Barab\'asi and R.~Albert, {\em
  Science} \textbf{286}, 509 (1999).

\bibitem{strogatz}
D.~J. Watts and S.~H. Strogatz,  {\em Nature} \textbf{393}, 440 (1998).

\bibitem{Girvan}
M.~Girvan and M.~E.~J. Newman,  {\em Proc. Natl. Acad. Sci. U.S.A.} \textbf{99}, 7821 (2002).

\bibitem{Hebert2011_prl}
L.~H\'ebert-Dufresne, A.~Allard, V.~Marceau, P.-A. No\"{e}l, and L.~J. Dub\'e,
   {\em Phys. Rev. Lett.} \textbf{107}, 158702 (2011).

\bibitem{ravasz}
E.~Ravasz, A.~L. Somera, D.~A. Mongru, Z.~N. Oltvai, and A.-L. Barab\'asi,
  {\em
  Science} \textbf{297}, 1551 (2002).

\bibitem{clauset}
A.~Clauset, C.~Moore, and M.~E.~J. Newman,  {\em Nature} \textbf{453},
  98 (2008).

\bibitem{Song05_nat}
C.~Song, S.~Havlin, and H.~A. Makse, 
  {\em Nature} \textbf{433}, 392 (2005).

\bibitem{Song06_natphys}
C.~Song, S.~Havlin, and H.~A. Makse,  {\em Nature Physics} \textbf{2}, 275 (2006).

\bibitem{boguna_natphys}
M.~Bogun\'{a}, D.~Krioukov, and K.~C. Claffy, {\em Nature Physics} \textbf{5}, 74 (2009).

\bibitem{boguna_natcomm}
M.~Bogun\'{a}, F.~Papadopoulos, and D.~Krioukov,  {\em Nature Communications} \textbf{1}, 1 (2010).

\bibitem{papadopoulos}
F.~Papadopoulos, M.~Kitsak, M.~A. Serrano, M.~Bogun\'{a}, and D.~Krioukov,
   {\em Nature} \textbf{489},
  537 (2012).

\bibitem{ravasz2}
E.~Ravasz and A.-L. Barab\'{a}si,  {\em Phys. Rev. E} \textbf{67}, 026112 (2003).

\bibitem{palla_pnas}
G.~Palla, L.~Lov\'{a}sz, and T.~Vicsek, 
  {\em Proc. Natl. Acad. Sci. U.S.A.} \textbf{107}, 7640 (2010).

\bibitem{lhd12}
L.~H\'ebert-Dufresne, A.~Allard, V.~Marceau, P.-A. No\"{e}l, and L.~J.
  Dub\'{e},  {\em Phys.
  Rev. E} \textbf{85}, 026108 (2012).
  
\bibitem{redner00}
P.~L.~Krapivsky, S.~Redner, and F.~Leyvraz,  {\em Phys.
  Rev. E} \textbf{85}, 4629 (2000).
  
\bibitem{doro00}
S.~N.~Dorogovtsev, J.~F.~F.~Mendes, and A.N.~Samukhin,  {\em Phys.
  Rev. E} \textbf{85}, 4633 (2000).
  
\bibitem{doro01}
S.~N.~Dorogovtsev, and J.~F.~F.~Mendes,  {\em Phys.
  Rev. E} \textbf{63}, 056125 (2001).
  
\bibitem{redner01}
P.~L.~Krapivsky, and S.~Redner  {\em Phys.
  Rev. E} \textbf{63}, 066123 (2001).
  
\bibitem{volz04}
E.~Volz  {\em Phys.
  Rev. E} \textbf{70}, 056115 (2004).
  
\bibitem{bianconi14}
G.~Bianconi, R.~K.~Darst, J.~Iacovacci, and S.~Fortunato {\em Phys.
  Rev. E} \textbf{90}, 042806 (2014).
  
\bibitem{redner05}
P.~L.~Krapivsky, and S.~Redner  {\em Phys.
  Rev. E} \textbf{71}, 036118 (2005).
  
\bibitem{redner13}
A.~Gabel, P.~L.~Krapivsky, and S.~Redner  {\em Phys.
  Rev. E} \textbf{88}, 050802(R) (2013).
  
  
\bibitem{Yule2}
G.~U. Yule,  {\em Philosophical Transactions of the Royal
  Society of London B} \textbf{213}, 21 (1925).

\bibitem{Gibrat}
R.~Gibrat, PhD thesis, Universit\'e de Lyon, 1931.

\bibitem{Simon}
H.~A. Simon, {\em Models of {M}an} (John Wiley \& Sons, New York, 1961).

\bibitem{Price}
D.~{de Solla Price}, {\em Journal of the American Society for Information
  Science} \textbf{27}, 292 (1976).

\bibitem{Zipf}
G.~K. Zipf, {\em Human {B}ehavior and the {P}rinciple of {L}east {E}ffort} (Addison-Wesley Press, Cambridge, 1949).

\bibitem{LHD13_HRN}
L.~H\'{e}bert-Dufresne, A.~Allard, J.-G. Young, and L.~J. Dub\'{e},
  {\em
  Phys. Rev. E} \textbf{88}, 062820 (2013).

\bibitem{Newman02_PhysRevLett}
M.~E.~J. Newman,  {\em Phys. Rev. Lett.} \textbf{89}, 208701 (2002).

\bibitem{GR}
I.~S. Gradshteyn and I.~M. Ryzhik, {\em {T}able of {I}ntegrals, {S}eries and {P}roducts}.
\newblock 7th ed. (Academic Press, Amsterdam, 2007).
\end{thebibliography}
\end{document}